\documentclass[a4paper]{article}

\usepackage{INTERSPEECH2019}
\usepackage{adjustbox}
\usepackage{enumitem}
\usepackage{hyperref}

\usepackage[labelformat=simple]{subcaption}

\usepackage[font=small,labelfont=bf]{caption}

\usepackage{pifont}

\title{The JHU Multi-Microphone Multi-Speaker ASR System\\for the CHiME-6 Challenge}
\name{$^{\ast}$Ashish Arora, $^{\ast}$Desh Raj, $^{\ast}$Aswin Shanmugam Subramanian, $^{\ast}$Ke Li, Bar Ben-Yair,\\ Matthew Maciejewski, Piotr \.Zelasko, Paola Garc\'ia, Shinji Watanabe, Sanjeev Khudanpur
\thanks{$^{\ast}$ equal contribution.}}
\address{
  Center for Language and Speech Processing \& Human Language Technology Center of Excellence \\ 
  The Johns Hopkins University, Baltimore, MD 21218, USA.
}
\email{\{aarora8,aswin,kli26\}@jhu.edu, draj@cs.jhu.edu}

\begin{document}

\maketitle

\begin{abstract}
This paper summarizes the JHU team's efforts in tracks 1 and 2 of the CHiME-6 challenge for distant multi-microphone conversational speech diarization and recognition in everyday home environments. 
We explore multi-array processing techniques at each stage of the pipeline, such as multi-array guided source separation (GSS) for enhancement and acoustic model training data, posterior fusion for speech activity detection, PLDA score fusion for diarization, and lattice combination for automatic speech recognition (ASR). We also report results with different acoustic model architectures, and integrate other techniques such as online multi-channel weighted prediction error (WPE) dereverberation and variational Bayes-hidden Markov model (VB-HMM) based overlap assignment to deal with reverberation and overlapping speakers, respectively. As a result of these efforts, our ASR systems achieve a word error rate of 40.5\% and 67.5\% on tracks 1 and 2, respectively, on the evaluation set.  This is an improvement of 10.8\% and 10.4\% absolute, over the challenge baselines for the respective tracks.
\end{abstract}

\noindent\textbf{Index Terms}: CHiME-6 challenge, robust speech recognition, speaker diarization, multi-channel, multi-speaker

\section{Introduction}

Far-field automatic speech recognition (ASR) and speaker diarization are important areas of research and have many real-world applications such as transcribing meetings~\cite{Hain2012TranscribingMW,Hori2012LowLatencyRM,Renals2017DistantSR,Yoshioka2018RecognizingOS} and in-home conversations~\cite{barker2018fifth}. Although deep learning methods (including end-to-end approaches) have achieved promising results for several tasks such as Switchboard~\cite{Xiong2016AchievingHP,wang2019espresso} and LibriSpeech~\cite{Lscher2019RWTHAS, Synnaeve2019EndtoendAF}, their performance remains unsatisfactory for far-field conditions in real environments, such as the CHiME-5 dataset~\cite{barker2018fifth}. This can be attributed to: (i) noise and reverberation in the acoustic conditions, (ii) conversational speech and speaker overlaps, and (iii) challenge-specific restrictions such as insufficient training data. 

Several advances have been made in the last decade to tackle the challenges offered by real, far-field speech. For ASR, this improvement can be attributed to improved neural network architectures~\cite{zorilatoshiba,kanda2018hitachi}, effective data augmentation techniques~\cite{zorila2019investigation}, and advances in speech enhancement~\cite{boeddeker2018front}. Previous work has tried tackling reverberation and noise present in the far-field recording by multi-style training with data augmentation via room impulse responses and background noises~\cite{ko2017study,wang2019jhu}. Recently, spectral augmentation has been successfully used for both end-to-end~\cite{park2019specaugment} and hybrid ASR systems~\cite{Zhou2020TheRA}. Adapting the acoustic model to the environment~\cite{seltzer2013investigation} and speaker~\cite{saon2013speaker} has also been studied. Another popular direction is front-end based approaches such as dereverberation~\cite{drude2018nara} and denoising through beamforming~\cite{anguera2007acoustic,Nakatani2019AUC}, which utilize multi-microphone data. 
% Dereverberation reduces the effect of inference introduced due to reflection introduced from multiple sources. In beamforming, the effect of noise and overlap is reduced by utilizing the spatial information. 
Far-field speaker diarization~\cite{Garca2019SpeakerDI} has also benefited from enhancement methods~\cite{Sun2020ProgressiveMN,Kataria2020FeatureEW} and approaches to handle overlapping speech~\cite{bullock2019overlap}. Recently, guided source separation (GSS)~\cite{boeddeker2018front} was proposed, which makes use of additional information such as time and speaker annotations for mask estimation. However, it requires a strong diarization system to perform good separation.  

In this paper, we describe a multi-microphone multi-speaker ASR system developed using many of these methods for the CHiME-6 challenge \cite{watanabe2020chime}. The challenge aims to improve speech recognition and speaker diarization for far-field conversational speech in challenging environments in a multi-microphone setting. The CHiME-6 data \cite{barker2018fifth} contains a total of 20 4-speaker dinner party recordings. Each dinner party is two to three hours long and is recorded simultaneously on the participants' ear-worn microphone and six microphone arrays placed in the kitchen, dining room, and the living room. The challenge consists of two tracks. Track 1 allows the use of oracle start and end times of each utterance, and speaker labels for each segment. This track focuses on core ASR techniques, and measures system performance in terms of transcription accuracy. Track 2 is a ``diarization+ASR'' track. It additionally requires end-pointing speech segments in the recording, and assigning them speaker labels, i.e diarization.  To this end, VoxCeleb2 data~\cite{nagrani2017voxceleb} is permitted for training a diarization system, and concatenated minimum-permutation word error rate (cpWER) is used to measure speaker-attributed transcription accuracy. %Our team participated in both tracks of the challenge.

\begin{figure}[t]
\centering
\includegraphics[width=0.7\linewidth]{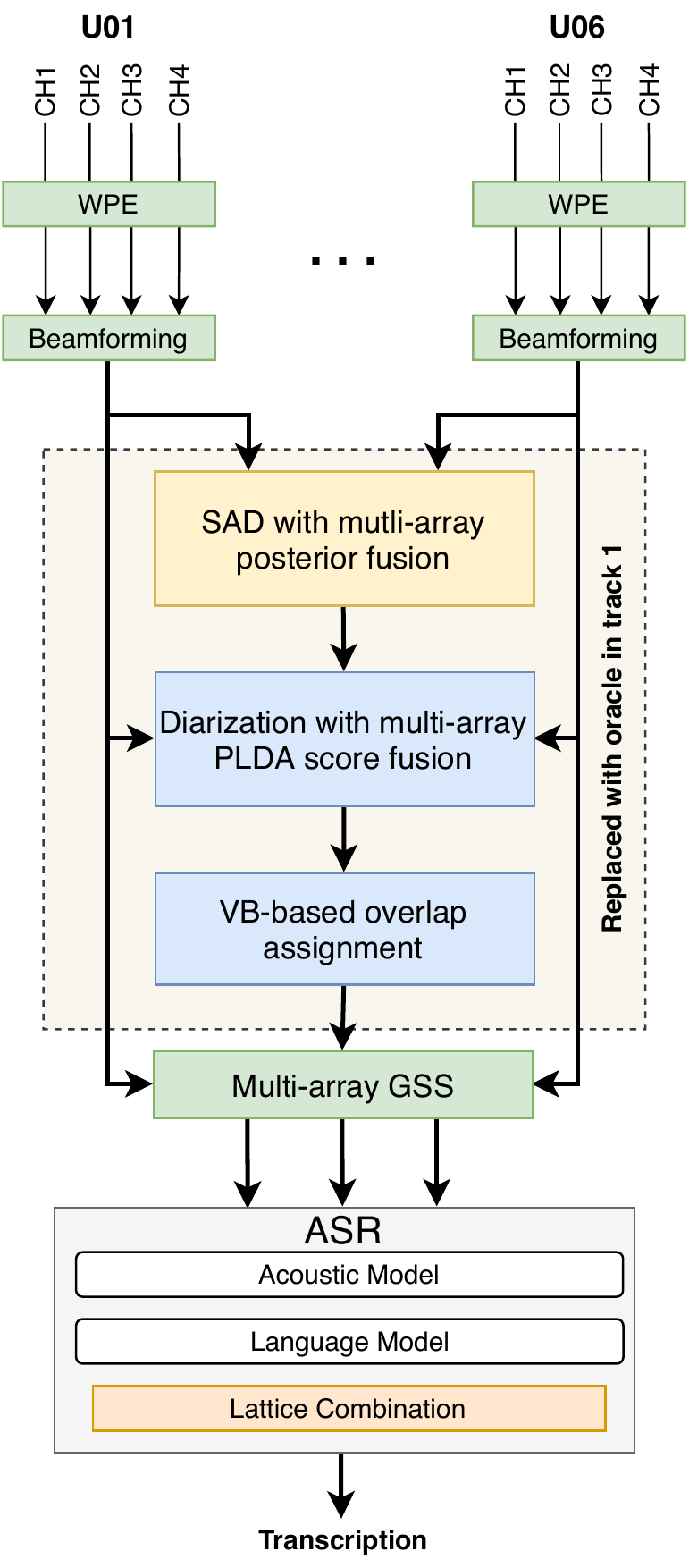}
\caption{Overview of the decoding pipeline for track 2. For track 1, we use a similar system, with the exception that the diarization module (shown in the dotted box in the figure) is replaced with oracle speech segments and speaker labels.}
\label{fig:pipeline}
\end{figure}

Our system for Track 2, as shown in Figure \ref{fig:pipeline}, consists of three main modules: \emph{enhancement}, \emph{diarization} and \emph{recognition}, described in Sections \ref{sec:enhancement}, \ref{sec:diarization} and \ref{sec:recognition} below. The \emph{enhancement module} performs (i) dereverberation using online multichannel weighted prediction error (WPE), followed by (ii) denoising with a weighted delay-and-sum beamformer, and (iii) multi-array guided source separation (GSS), described below. In the \emph{diarization module}, the beamformer outputs, one per array, are used for (i) speech activity detection and (ii) speaker diarization, both of which fuse information across arrays to improve accuracy, and (iii) overlap-aware variational Bayes hidden Markov model (VB-HMM) resegmentation to assign multiple speakers to overlapped speech regions.  Speaker-marks from this diarization module are used in the multi-array GSS (part of the enhancement module) to produce enhanced, speaker-separated waveforms.  The \emph{recognition module} processes the GSS output using (i) an acoustic and n-gram language model for ASR decoding, (ii) an RNN language model for lattice rescoring, and (iii) sMBR lattice combination.  We augment the clean acoustic training data with dereverberated, beamformed and GSS-enhanced far-field data to match the test conditions.

The diarization module is replaced with oracle speech segments and speaker labels in our system for Track 1.

\section{Speech Enhancement}
\label{sec:enhancement}

\subsection{Dereverberation and Denoising}
\label{sec:wpe+beamformit}

We used an online version of the publicly available NARA-WPE~\cite{drude2018nara} implementation of weighted prediction error (WPE) based dereverberation for multi-channel signals~\cite{nakatani2010speech} for all the channels in each array. This was followed by array-level weighted delay-and-sum beamforming using the BeamformIt tool~\cite{anguera2007acoustic}. All further processing was done on the dereverberated and beamformed signals.

\subsection{Guided Source Separation (GSS)}
\label{sec:gss}

Multi-array GSS~\cite{boeddeker2018front,kanda2019guided} was applied to enhance target speaker speech signals. For track 1, we used oracle speech segmentations and speaker labels, while for track 2, we used the segmentation estimated by the speaker diarization module described in Section \ref{sec:diarization}. In GSS, the
source activity pattern for each speaker derived from the segmentation aids in resolving the permutation ambiguity. A context window is used to obtain an extended segment which can bring sufficient sparsity in the activity pattern of the target speaker to further reduce speaker permutation issues. We found a 20-second context to be ideal from our experiments. GSS gave a relative WER improvement of 9.3\% and 6.5\% on Dev and Eval respectively for track 2 over the baseline delay-and-sum beamformer, as seen in Section \ref{sec:chime-results}, Table \ref{tab:asr_results2}.

\section{Speaker Diarization}
\label{sec:diarization}

\begin{figure*}[t]
\centering
\includegraphics[width=0.8\linewidth]{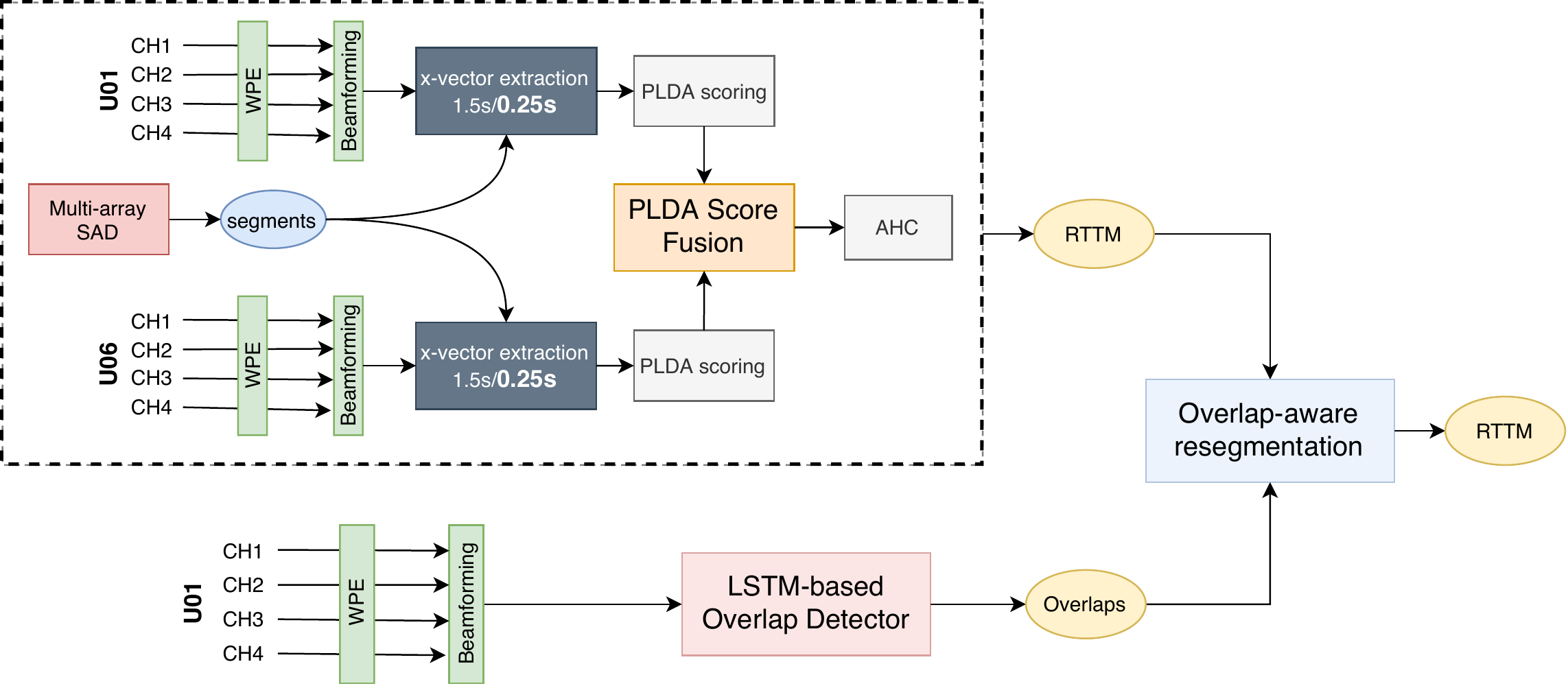}
\caption{The two-pass speaker diarization module.  Synchronous SAD marks across microphone arrays enables PLDA score fusion before AHC in first-pass diarization (dotted rectangle). Overlap detection (bottom) enables the second-pass resegmentation, initialized by the first-pass output (upper RTTM), to assign more than one speaker to overlapped-speech regions in the final output (right RTTM).}
\label{fig:diarization}
\end{figure*}

\subsection{Speech Activity Detection}
\label{sec:SAD}

For speech activity detection (SAD), we first trained a neural network classifier to assign each frame in an utterance a label from $\cal C$ = \{\textit{silence}, \textit{speech}, \textit{garbage}\}. We used the architecture shown in Table~\ref{tab:sad_dnn}, consisting of time-delay neural network (TDNN) layers to capture long temporal contexts \cite{Peddinti2015ATD}, interleaved with stats-pooling layers to aggregate utterance-level statistics. To obtain classifier training targets, we used a speaker-independent GMM-HMM ASR system to align whole recordings with the training transcriptions.
To leverage multiple channels, we carried out posterior fusion of the classifier outputs across all the arrays at test time. i.e. if $p^k_t(i)$ denotes the classifier's probability for class $i\in C$ at frame $t$ based on array $k$, then $p_t(i) = f(p^k_t(i))$, where $f$ is the fusion criterion.

We then post-processed the per-frame classifier outputs to enforce minimum and maximum speech/silence durations, by constructing a simple HMM whose state transition diagram encodes these constraints, treating the per-frame SAD posteriors $p_t(\cdot)$ like emission probabilities, and performing Viterbi decoding to obtain the most likely SAD label-sequence.

\begin{table}[tb]
\centering
\caption{Neural network architecture for speech activity detection (SAD).  $T$ is the input length, $\cal C$ the set of output classes.}
\label{tab:sad_dnn}
\begin{adjustbox}{max width=\linewidth}
\begin{tabular}{cccc}
\toprule
\thead{\textbf{Layer}} & \thead{\textbf{Layer context}} & \thead{\textbf{Total}\\\textbf{context}} & \thead{\textbf{Input$\times$Output}} \\
\midrule
tdnn1 & [$t-2$, $t+2$] & 5 & $150\times 256$ \\
tdnn2 & [$t-1$, $t+2$] & 8 & $1024\times 256$ \\
tdnn3 & \{$t-3$, $t$, $t+3$, $t+6$\} & 17 & $1024\times 256$ \\
stats1 & [0, $T$) & $T$ & \\
tdnn4 & \{$t-6$, $t$, $t+6$, $t+12$\} & $T$ & $1024 \times 256$\\
stats2 & [0, $T$) & $T$ & \\
%tdnn5 & \specialcell{\{$t-12$, $t$, $t+12$,\\ $t+24$\}} & $T$ & 256 $\times C$ \\
tdnn5 & \{$t-12$, $t$, $t+12$, $t+24$\} & $T$ & 256 $\times |\cal C|$ \\
\bottomrule
\end{tabular}
\end{adjustbox}
\end{table}

\begin{table}[t]
\centering
\caption{Reducing SAD errors by fusing posterior probabilities from multiple arrays. SAD errors are comprised of missed speech (MS) and false alarms (FA).}
\label{tab:sad_results}
\begin{adjustbox}{max width=\linewidth}
\begin{tabular}{lcccccc}
\toprule
\multicolumn{1}{c}{\multirow{2}{*}{\textbf{System}}} & \multicolumn{3}{c}{\textbf{Dev}}           & \multicolumn{3}{c}{\textbf{Eval}}          \\
                                 & \textbf{MS} & \textbf{FA} & \textbf{Total} & \textbf{MS} & \textbf{FA} & \textbf{Total} \\
\midrule
Baseline (U06) & 2.7 & \textbf{0.6} & 3.3 & 4.4 & \textbf{1.5} & 5.9 \\
Posterior Mean & 1.7 & 0.7 & 2.4 & 3.2 & 1.9 & \textbf{5.1} \\
Posterior Max & \textbf{1.1} & 0.8 & \textbf{1.9} & \textbf{2.4} & 2.8 & 5.2 \\
\bottomrule
\end{tabular}
\end{adjustbox}
\end{table}

For the fusion criterion, we experimented with simple element-wise mean and max, and found them to be effective. From Table~\ref{tab:sad_results}, we can see that applying posterior fusion across arrays improved the SAD error by approximately 34\% relative. Since most of the gain comes from reduction in missed speech, this also positively impacts downstream recognition rate.

We also tried microphone-level posterior fusion jointly across all the arrays, but it did not yield any improvements over using the beamformed signals as described above.
 
\subsection{First Pass Speaker Diarization}
\label{sec:xvec-diarization}

%For diarization, we performed agglomerative hierarchical clustering (AHC) on the probabilistic linear discriminant analysis (PLDA) similarity scores computed between x-vectors~\cite{sell2018diarization} extracted from 1.5s windows with 0.25s stride~\cite{landini2019but}. The x-vector extractor consists of TDNN layers with stats pooling (similar to \cite{snyder2018x}), and was trained on VoxCeleb data~\cite{nagrani2017voxceleb} augmented with CHiME-6 background noises and simulated room impulse responses (RIRs)~\cite{ko2017study}. The PLDA parameters were trained on a 100k subset of the CHiME-6 training data. Similar to our posterior fusion method in SAD, we perform multi-array PLDA score fusion before the clustering stage. Again, using element-wise max for the fusion was found to be effective.

Our first-pass diarization followed the method described in \cite{sell2018diarization}.  The test recordings were cut into into overlapping 1.5 second \emph{segments} with a 0.25 second\footnote{The CHiME baseline system used a 0.75 second stride.} stride \cite{landini2019but}, an x-vector was extracted from each segment, and agglomerative hierarchical clustering (AHC) was performed on the x-vectors using the probabilistic linear discriminant analysis (PLDA) score of each pair of x-vectors as their pairwise similarity.

The x-vector extractor we used is similar to that of \cite{snyder2018x}: it is comprised of several TDNN layers and stats-pooling, and was trained, per Challenge stipulations, on only the VoxCeleb data \cite{nagrani2017voxceleb}.  Data augmentation was performed by convolving the audio with simulated room impulse responses \cite{ko2017study} and by adding background noises from the CHiME-6 training data.  PLDA parameters were trained on (x-vectors of) 100k speech segments from the ca 40 speakers in the CHiME-6 training data.

Similar to the SAD posterior probability fusion described in Section \ref{sec:SAD}, we investigated improving diarization by leveraging multiple microphone( array)s at test time.  To compute the pairwise similarity of two 1.5 second segments during clustering, we fused the PLDA scores for their x-vectors extracted from different arrays.  We found that multi-array PLDA score fusion, specifically the element-wise maximum across arrays, provided noticeable gains.

\subsection{Overlap-Aware Resegmentation}
\label{sec:vb-reseg}

Since AHC is not designed to handle overlapping speakers, we resegmented the audio using an \emph{overlap-aware} version of the VB-HMM of \cite{Dez2018SpeakerDB}.  Speaker labels from the first-stage diarization of Section \ref{sec:xvec-diarization} were used to initialize the per-frame speaker posterior matrix, also known as the $\boldsymbol{Q}$-matrix, and one iteration of VB-HMM inference was performed to convert this (binary) $\boldsymbol{Q}$-matrix into per-frame speaker-probabilities.  Separately, we trained an \emph{overlap detector}|a 2-layer bidirectional LSTM with SincNet input features \cite{Ravanelli2018SpeakerRF} and binary (non-overlapped/overlapped speech) output labels|using the CHiME-6 training data.  The per-frame decisions of the overlap detector on the test data were then used to assign each frame to either one or two most likely speakers according to $\boldsymbol{Q}$, as described in \cite{bullock2019overlap}.  An unintended attribute of VB resegmentation was a significant number of very short segments.  The computational complexity of the GSS module (cf Section \ref{sec:gss}) is severely impacted by this growth in the number of segments.  We therefore removed all segments shorter than 200ms.

The overall diarization process is shown in Figure \ref{fig:diarization}.

\subsection{Diarization Performance}

The CHiME Challenge provided two ``ground truths'' for diarization, i.e. two NIST-style rich transcription time marks (RTTM): one based on utterance-level time marks by human annotators (named Annotation RTTM), and another on forced-alignment of the acoustics to the transcripts (resp Alignment RTTM).  The diarization output was scored against each RTTM using the DiHARD \texttt{dscore} toolkit\footnote{https://github.com/nryant/dscore}, and diarization error rate (DER) as well as Jaccard error rate (JER) were computed.

\begin{table}
\centering
\caption{Diarization performance on track 2, showing the impact of the modifications of Sections \ref{sec:xvec-diarization} and \ref{sec:vb-reseg} to the baseline x-vector/AHC system. Some of the improvement derives from the improved SAD of Section \ref{sec:SAD}.}
\label{tab:diar_results}
\begin{adjustbox}{max width=\linewidth}
\begin{tabular}{lcccc}
\toprule
\multicolumn{1}{c}{\multirow{2}{*}{\textbf{System}}} & \multicolumn{2}{c}{\textbf{Dev}} & \multicolumn{2}{c}{\textbf{Eval}} \\
\multicolumn{1}{c}{}                                 & \textbf{DER}    & \textbf{JER}   & \textbf{DER}    & \textbf{JER}    \\
\midrule
\multicolumn{5}{c}{\textit{Alignment RTTM}} \\
\midrule
Baseline (U06)  & 63.42  & 70.83 & \textbf{68.20} & 72.54  \\
PLDA Fusion & 63.97 & 71.65 & 71.56 & 71.32 \\
+ 0.25s stride   & 61.00 & 66.23 & 69.64 & 69.81 \\
+ overlap assign. & \textbf{58.18} & \textbf{59.92} & 69.92 & \textbf{65.64}  \\
\midrule
\multicolumn{5}{c}{\textit{Annotation RTTM}} \\
\midrule
Baseline (U06)  & 61.62  & 69.84 & 62.01 & 71.43  \\
PLDA Fusion & 60.09 & 70.31 & 62.97 & 70.09 \\
+ 0.25s stride   & 57.85 & 65.36 & 61.60 & 69.35 \\
+ overlap assign. & \textbf{50.43} & \textbf{57.81} & \textbf{58.26} & \textbf{64.38}  \\
\bottomrule
\end{tabular}
\end{adjustbox}
\end{table}

Table~\ref{tab:diar_results} shows improvements in diarization performance, relative to the AHC baseline, due to PLDA score fusion, using a 0.25s stride (v/s 0.75s), and overlap-aware resegmentation.  The gains from PLDA fusion across arrays appears modest and somewhat inconsistent relative to the single array (U06) DER, but was consistently better than the \emph{average} single-array DER across the six arrays.  The shorter (0.25s) x-vector stride yielded a robust improvement, and the most significant improvement came from overlap detection and multiple-speaker assignment (denoted overlap assign.)

Finally, note that diarization performance seems to degrade (particularly for the Eval data) when scored against the Alignment RTTM, but shows significant improvement with the Annotation RTTM.  The former stipulates tighter speech boundaries by, for instance, marking short pauses between words as non-speech.  This increases the (measured) false alarm errors of our diarization module.  However, retaining such pauses is beneficial for the downstream ASR task, by inducing more appropriate utterance segmentation.

\section{Automatic Speech Recognition}
\label{sec:recognition}
We built a hybrid DNN-HMM system using algorithms and tools available in the \href{https://github.com/kaldi-asr/kaldi}{Kaldi} ASR toolkit.

\subsection{Acoustic Modeling}

Our baseline acoustic model (AM) was a 15-layer factorized time-delay neural network or TDNN-F \cite{PoveyEtAlIS2018}. Training data for this model comprised of 80h clean CHiME-6 audio from the 2 worn-microphones of the speaker of each transcribed utterance, 320h from a 4x distortion of this clean audio using synthetic room impulse responses and CHiME-6 background noises, and 200h raw far-field audio derived by randomly sampling utterances from the many arrays.  These 600h were subject to 0.9x and 1.1x speed-perturbation to yield 1800h of AM training data.  The baseline TDNN-F was trained with the lattice-free maximum mutual information (LF-MMI) objective.  A context-dependent triphone HMM-GMM system was first trained using standard procedures in Kaldi, and frame-level forced-alignments were generated between the speech and reference transcripts to \emph{guide} LF-MMI training \cite{PoveyEtAlIS2016}.

Our CHiME-6 Challenge baseline system used this acoustic model and achieved a word error rate (WER) of 51.8\% and 51.3\% respectively on the CHiME-6 Dev and Eval test sets.  We will designate this AM by (a) in this paper.

We then experimented with several other model architectures and data selection and augmentation methods.

%We incrementally trained monophone, triphone (tri1), LDA-MLLT (tri2), and SAT (tri3) HMM-GMM systems\footnote{Names in parentheses are \href{https://github.com/kaldi-asr/kaldi}{Kaldi} notations for these models.}, and used the tri3 model for generating alignments for neural network training. Before tri3 training, we also re-computed the pronunciation and silence probabilities using the GMM tri2 system.

%We used the architecture (CNN + TDNN-F) shown in Fig.~\ref{fig:acoustic}, consisting of a convolution neural network (CNN) to learn 2-D linear filters and factorized time-delay neural network (TDNN-F)~\cite{povey2018semi} to have deep layers with long temporal context. It consists of 6 CNN and 16 TDNN-F layers with ResNet-style skip connections. We use batch normalization \cite{ioffe2015batch} and L2-regularization on all CNN and TDNN-F layers. During training, online i-vectors were extracted every ten frames using an extractor trained on the same training data as the acoustic model. We also appended oracle overlap information to the acoustic features, as described in Section  \ref{overlap_info}. At inference time, we used multi-array GSS enhancement, similar to the baseline system. We performed 2-stage decoding~\cite{peddinti2015jhu}, where utterance-level i-vectors are used in the first pass, and then reweighted only with high confidence regions in the second pass.

\begin{figure}[t]
\centering
\includegraphics[width=0.7\linewidth]{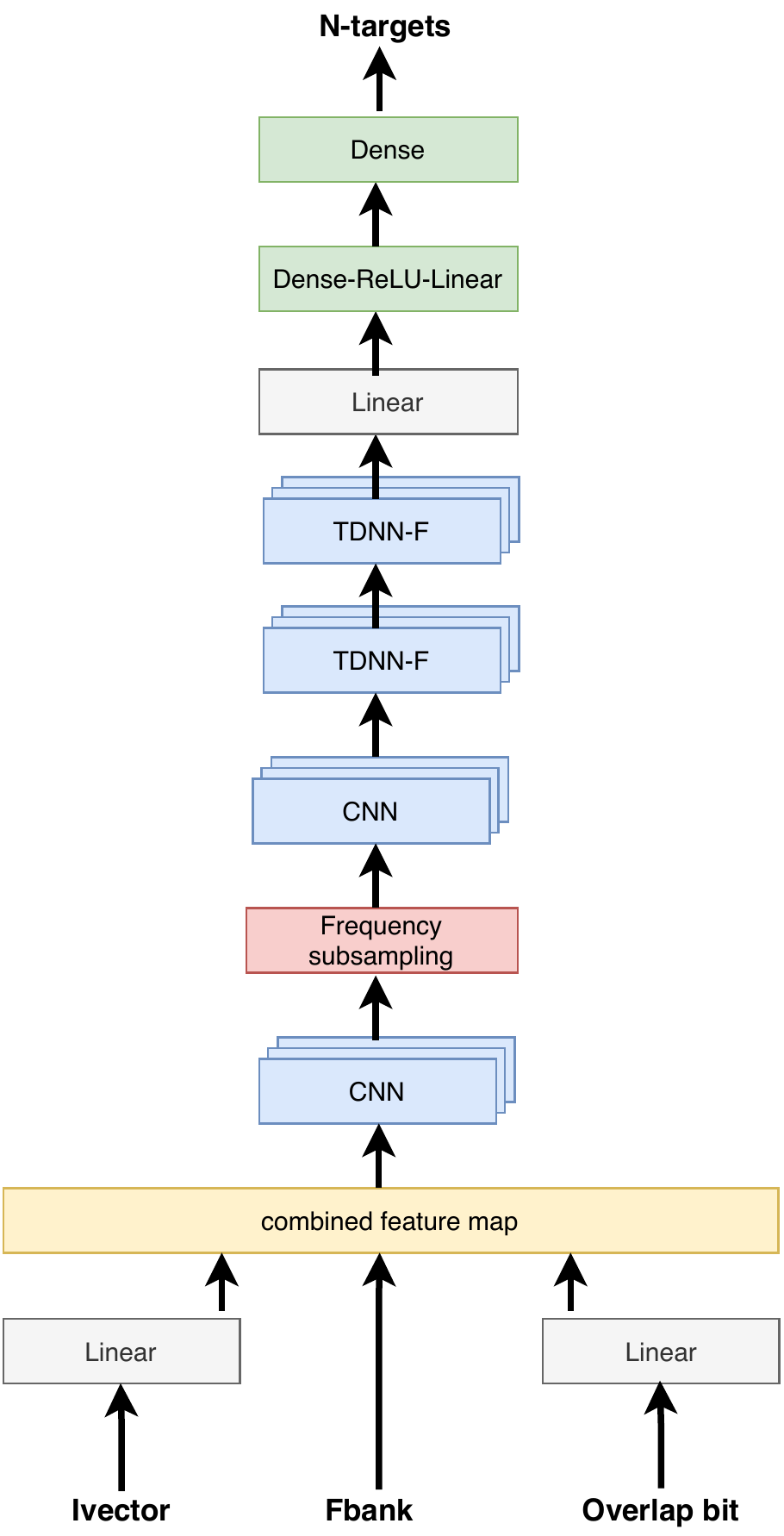}
\caption{Illustration of the acoustic model architecture, inputs and outputs.  N-targets is the number of leaves (context dependent HMM states) in the bi-phone clustering trees.}
\label{fig:acoustic}
\end{figure}

\subsubsection{Neural Network Architecture}
\label{sec:arch}

%In addition to the 15-layer TDNN-F in the baseline setup, we explored several neural network architectures for acoustic model training, as shown in Table~\ref{tab:architecture}. These acoustic models were trained on 500h of a combination of worn mic utterances (80h), reverberated and noise augmented close-talk data (320h), multi-array GSS enhanced data (40h), and raw far-field data (60h), while the TDNN-F setup was trained on 1800h of data as mentioned earlier in Section~\ref{data_augmentation}.
We first trained an AM with 6 CNN layers comprising 3$\times$3 convolutional kernels, and 16 TDNN-F layers using the 1800h training set, designated model (b) in this paper, and lowered the WERs to 49.6\% and 49.3\% on the Dev and Eval sets, respectively.  But we discovered that these models take an inordinate amount of time to train, limiting our ability for exploration.

To expedite the turnaround time of AM training experiments, and informed by experiments in data selection and augmentation described in Section \ref{data_augmentation}, we created a 500h training set comprised of the 80h of clean ear-worn microphone audio and 320h of 4x distortions of the clean audio using synthetic room impulse responses and CHiME-6 background noises, as above, supplemented with 60h raw far-field audio derived by randomly sampling utterances from the many arrays and 40h from multi-array GSS enhancement, as described in Section \ref{sec:gss}.

With this 500h training set, we experimented with the following AM architectures.

\begin{enumerate}[wide, labelwidth=!, labelindent=0pt, label=(\alph*), start=3]

\item
\textbf{6 CNN + 12 TDNN-F}, comprised of 6 CNN layers with 3x3 convolutional kernels, followed by 12 TDNN-F layers. Frequency subsampling by a factor of two was applied at the third and fifth convolutional layers.  The TDNN-F layers were 1536-dimensional with a factorized (lower) rank of 160.

\item
\textbf{6 CNN + 10 TDNN-F + 2 Self-attention}, in which the last two TDNN-F layers were replaced by self-attention layers. Each self-attention layer had 15 attention heads with key and value dimensions of 40 and 80, respectively.

\item
\textbf{CNN + TDNN + LSTM}, comprised of 6 CNN layers, followed by 9 TDNN and 3 LSTM layers, interleaved (3+1)$\times$3.  The CNN layers the are same as above, TDNN layers have 1024-dimensional hidden units, and the LSTM has a 1024-dim layer and 256-dim output and recurrent projections.

\end{enumerate}

The input to these AMs was 40-dim log-Mel filterbank coefficients, and the outputs were context-dependent HMM states (Senones) derived from a left-biphone tree.  We determined empirically that a tree with 4500 leaves, an L$_2$-regularization value of 0.03 for the CNN and TDNN-F layers, and 0.04 for the dense layers performed well.  Using  6-epochs instead of 4-epochs (in the baseline) further improved the AMs.
  
As shown in Table~\ref{tab:architecture}, the CNN + TDNN-F architecture performed better than the others, so we selected it for subsequent experiments.

\begin{table}
\centering
\caption{Track 1 ASR WERs for AM architecture described in Section \ref{sec:arch}.  The CNN + TDNN-F configuration works best.}
\label{tab:architecture}
\begin{adjustbox}{max width=\linewidth}
\begin{tabular}{lcc}
\toprule
\textbf{(Model) Architecture} & \textbf{Dev} & \textbf{Eval} \\
\midrule
(a) 15 TDNN-F & 51.8\% & 51.3\% \\
(b) 6 CNN + 16 TDNN-F & 49.6\% & 49.3\% \\
(c) 6 CNN + 12 TDNN-F & \textbf{48.3}\% & \textbf{48.5}\%  \\
(d) 6 CNN + 10 TDNN-F + 2 SA & 49.9\% & 49.4\%  \\
(e) 6 CNN + 3 $\times$ (3 TDNN + 1 LSTM) & 50.1\%& 49.8\%  \\
\bottomrule
\end{tabular}
\end{adjustbox}
\end{table}

\subsubsection{Training Data Selection and Augmentation}
\label{data_augmentation}

%We used CNN + TDNN-F model architecture, and multi-array GSS based enhancement for the development and evaluation data, in our experiments in  Table~\ref{tab:data}. In the baseline setup, the training data consisted of around 1800 hours of worn microphone data(80h), raw far-field data (200h), and worn microphone data augmented with reverberation and noises extracted from the training data (320h), with the whole data augmented with speed perturbation.

%In order to reduce the mismatch between the training data and the development data, we augmented the training data with additional enhancend copies obtained using multi-array GSS and weighted delay-and-sum beamforming. From the original 40h of data, we obtained a 280h combination of worn mic utterances (80h), beamformed array data (160h), and multi-array GSS enhanced data (40h). Cleanup and speed perturbation were performed on the dataset, finally resulting in 675h of training data. Similar to ~\cite{zorila2019investigation}, we also found that using enhanced far-field data can improve the acoustic model performance and can reduce the amount of data required to train the model. Speed-perturbation augmentation was also found to improve the performance of the acoustic model by around 2\% for this setup.

\begin{table}[t]
\centering
\caption{Track 1 WERs when training the 6 CNN + 16 TDNN-F AM on only enhanced (and not reverberated, noisy) speech.}
\label{tab:data}
\begin{adjustbox}{max width=\linewidth}
\begin{tabular}{lcc}
\toprule
\textbf{(Model) Training Data}& \textbf{Dev} & \textbf{Eval} \\
\midrule
(b) 1800h incl. reverb'ed \& raw far-field& 49.6\% & 49.3\%  \\
(f)  675h incl. only clean \& enhanced& \textbf{44.5\%} & \textbf{44.9\%}  \\
(g)  675h of model (f) + sp/sil probs& 45.0\% & 45.3\%  \\
(h)  3 $\times$ 500h from models (c)-(e)& 44.6\% & 45.4\%  \\
\bottomrule
\end{tabular}
\end{adjustbox}
\end{table}

Inspired by the findings of \cite{zorila2019investigation}, where they reported good performance on CHiME data not by using the matched far-field speech, or synthetically reverberated speech, for AM training, but instead using speech enhancement during both AM training and test time, we created a new AM training data set as follows.  We applied beam-forming to the 4 microphone arrays to obtain 4 $\times$ 40h of speech, obtained another 40h from multi-array GSS described in Section \ref{sec:gss}, and combined them with the 80h of ear-worn microphone data described above. Data clean-up was applied to this 280h data set, to remove utterances that failed forced-alignment under a narrow beam, followed by speed perturbation, resulting in a 675h AM training data set.
% Speed perturbation was applied to this 280h data set, to grow it to 840h, followed by data clean-up to remove utterances that failed forced-alignment under a narrow beam, resulting in a 675h AM training data set.

Note from Table \ref{tab:data} that careful data selection indeed confirmed the findings of \cite{zorila2019investigation}, reducing the WER to 44.5\% and 44.9\% respectively on the Dev and Eval sets.  The model that attained this is designated (f) in this paper.

We also trained two additional AMs, one with the same 675h of data as the model (f), but with improved estimates of the inter-word silence and pronunciation probabilities (see \cite{ChenEtAlIS2015}), and another with speed perturbation of the 500h data set of models (c)-(e).  These models are designated (g) and (h) respectively in Table \ref{tab:data}.  While they do not perform better than the model (f), they were used for system combination in track 1, as described later in Table \ref{tab:asr_results} below.

\subsubsection{Overlap-Aware Training}
\label{overlap_info}

Since the CHiME data have a significant proportion of overlapped speech, we looked into providing the AM a 1-bit input indicating the presence/absence of overlap.  This \emph{overlap bit} was determined during training from the time alignments, and used alongside the 40-dim filter-bank features and 100-dim i-vectors.  We first projected the overlap bit to 40-dim, and the i-vector to 200-dim, and applied batch normalization and L$_2$ regularization.  We then combined the two with the filter-bank features to create a single-channel input to the first CNN layer of model (f) described above.

The resulting model is designated (i) in Table~\ref{tab:overlap}, which illustrates that knowledge of the presence of overlapped speech yields a modest WER improvement in track 1 conditions.
While we could have used model (i) in track 2 by using the output of the overlap detector of Section \ref{sec:vb-reseg} as the overlap bit, we did not have sufficient time to carefully conduct these experiments.

%\begin{table}[t]
%\centering
%\caption{WER on Dev and Eval set with overlap information. Both setups are trained with 6 %CNN and 16 TDNN-F layers neural network architecture.}
%\begin{tabular}{ccc}
%\toprule
%\textbf{Overlap bit} & \textbf{Dev (\%)} & \textbf{Eval (\%)} \\
%\midrule
% \xmark & 44.5 & 44.9  \\
% \cmark & \textbf{44.4} & \textbf{44.5}  \\
%\bottomrule
%\end{tabular}
%\end{table}

\begin{table}[t]
\centering
\caption{Track 1 WERs when the presence of overlapped speech is known to the acoustic model.}
\label{tab:overlap}
\begin{adjustbox}{max width=\linewidth}
\begin{tabular}{lcc}
\toprule
\textbf{(Model) Input features}& \textbf{Dev} & \textbf{Eval} \\
\midrule
(f) log-Mel filter-bank and i-vector & 44.5\% & 44.9\%  \\
(i) + overlapped speech indicator bit& \textbf{44.4}\% & \textbf{44.5}\%  \\
\bottomrule
\end{tabular}
\end{adjustbox}
\end{table}

\subsection{Language Modeling and Rescoring}

We used the training transcriptions to build our language models (LMs). We used a 3-gram LM trained with the SRILM toolkit~\cite{stolcke2002srilm} in the first pass decoding. For neural LMs, we used Kaldi to train recurrent neural network LMs (RNNLMs)~\cite{xu2018neural}. We performed a pruned lattice rescoring~\cite{xu2018pruned} with a forward and a backward (reversing the text at the sentence level) LSTM. We first rescored with the forward LSTM and then performed another rescoring on top of the rescored lattices using the backward LSTM. 
Both LSTMs are 2-layer projected LSTMs with hidden and projection layer dimensions 512 and 128, respectively. We also used L$_2$-regularization on the embedding, hidden, and output layers.

\subsection{Lattice Combination}

For track 1, we used the lattice combination method to combine four CNN + TDNN-F acoustic models, named models (f), (g), (h) and (i) in Section \ref{data_augmentation}, all with WER's in the 44\%-45\% range, as seen in Tables \ref{tab:data} and \ref{tab:overlap}.  While models (f), (g) and (i) were trained with only worn-microphone and enhanced far-field data, model (h) included raw far-field data and synthetically reverberated/noisy data.  For track 2, we used only one acoustic model (f), we performed GSS twice with two individual arrays and once with all arrays together, followed by lattice combination. Diarization output was shared by all input array signals, and minimum Bayes risk decoding \cite{xu2011minimum} was applied on top of the combined lattice.

\section{CHiME Challenge Results}
\label{sec:chime-results}

We show improvement in WER for tracks 1 and 2, obtained using the different modifications described in the previous sections, in Table~\ref{tab:asr_results} and Table~\ref{tab:asr_results2}, respectively. 

From Table~\ref{tab:asr_results}, we see that using a larger CNN + TDNN-F acoustic models improved performance over the TDNN-F baseline system by approx. 2\% absolute. For further improvement, we trained the AM on data closer to test conditions by augmenting the clean worn-microphone data with beamformed and GSS-enhanced data \cite{zorila2019investigation}. This provided almost 5\% absolute WER reduction. Adding the overlap bit provided a modest improvement. Lattice combination and LM rescoring were individually effective, and their combination provided a significant WER reduction of 4\% absolute.  Cumulatively, we obtained more than 10\% absolute WER improvement over the Challenge baseline system for track 1.

Table~\ref{tab:asr_results2} shows a similar step-wise WER improvement for track 2. Again, we obtained 2\% improvement using the larger CNN + TDNN-F architecture. Using multi-array fusion techniques in SAD and first-pass diarization reduced missed speech and speaker confusion, resulting in additional 4\% and 2\% improvement on the Dev and Eval sets, respectively. Using GSS and overlap-aware VB-HMM diarization provided significant improvements of 5\%--8\%, since they permit separating the overlapped speech for ASR. Finally, LM rescoring and lattice combination of ASR output from multiple input streams (but one ASR system) provided additional gains, similar to those observed in track 1 from combining multiple ASR systems.  Cumulatively, we obtained an absolute WER improvement of 17\% and 10\% respectively on the Dev and Eval sets.

\begin{table}[t]
\centering
\caption{Stepwise improvement in WER on Track 1.}
\label{tab:asr_results}
\begin{adjustbox}{max width=\linewidth}
\begin{tabular}{lcc}
\toprule
% \multicolumn{1}{c}{\multirow{2}{*}{\textbf{System}}} & \multicolumn{2}{c}{\textbf{Track 1}} & \multicolumn{2}{c}{\textbf{Track 2}} \\
%\multicolumn{1}{c}{} \\
\textbf{System components (acoustic model)}  &\textbf{Dev}   & \textbf{Eval}\\
\midrule
% Baseline (U06)                        & 51.75 & 51.29 & 84.33 & 78.08\\
% CNN-TDNN-F AM                         & 49.59 & 49.37 & 84.17 & 76.78\\
%+Augmentation                          & 44.60 & 45.40 & 81.63 & 75.98\\
Baseline TDNN-F (a)                            & 51.8\%        & 51.3\%\\
CNN-TDNN-F (b)                          & 49.6\%        & 49.4\%\\
+ Data selection and augmentation (f)    & 44.5\%        & 44.9\%\\
+ Overlap feature (i)                    & 44.4\%        & 44.5\%\\
% the commented rescoring results is on top of only +Augmentation
%+ Rescoring                            & 43.1\%        & 43.3\%\\
+ RNN LM rescoring (i)                   & 42.8\%        & 42.9\%\\
%+ Lattice combination                   & 41.9\%        & 42.4\%\\
+ Lattice combination (f)+(g)+(h)+(i)    & 41.8\%        & 42.1\%\\
% +Model combination and Rescoring      & 40.8\%        & 41.2\%\\
+ Lattice combination \& LM rescoring         & 40.3\%        & 40.5\%\\
% +Lattice combination and Rescoring    & 40.7\%        & 40.8\%\\
\bottomrule
\end{tabular}
\end{adjustbox}
\end{table}

\begin{table}
\centering
\caption{Stepwise improvement in WER on Track 2.}
\label{tab:asr_results2}
\begin{adjustbox}{max width=\linewidth}
\begin{tabular}{lcc}
\toprule
% \multicolumn{1}{c}{\multirow{2}{*}{\textbf{System}}} & \multicolumn{2}{c}{\textbf{Track 1}} & \multicolumn{2}{c}{\textbf{Track 2}} \\
%\multicolumn{1}{c}{} \\
\textbf{System components (acoustic model)}   &\textbf{Dev}   &\textbf{Eval}\\
\midrule
Baseline TDNN-F (a)                            & 84.3\%        & 77.9\%\\
CNN-TDNN-F (f)                          & 82.5\%        & 75.8\%\\
+ Multi-array SAD \& PLDA fusion (f)     & 78.3\%        & 73.6\%\\
+ Multi-array GSS (f)                    & 71.0\%        & 68.8\%\\
+ VB Overlap Assignment (f)              & 69.3\%        & 68.8\%\\
+ RNN LM Rescoring (f)                   & 68.7\%        & 67.9\%\\
+ Lattice combination (f)                & 68.3\%        & 68.3\% \\
+ Lattice combination \& LM rescoring (f)   & 67.8\%        & 67.5\%\\
\bottomrule
\end{tabular}
\end{adjustbox}
\end{table}

\section{Conclusion}

We described our system for the sixth CHiME challenge for distant multi-microphone conversational speaker diarization and speech recognition in everyday home environments. We explored several methods to incorporate multi-microphone and multi-array information for speech enhancement, diarization, and ASR. For track 1, most of the improvements in WER were obtained from data selection and augmentation, and language model rescoring. Through careful training data selection, we reduced the training time of the system 3-fold while also improving its performance. In track 2, array fusion and overlap handling in the diarization module provided more accurate speaker segments than the Challenge baseline, resulting in improved speech enhancement via multi-array GSS. The gains from acoustic modeling and RNNLM rescoring developed in track 1 also largely carried over to track 2.

\section{Acknowledgement}
We thank Daniel Povey for the tremendous support throughout the system development, Jing Shi for help with data simulation in our trials with neural beamforming, and Yiming Wang for help with the lattice combination setup. This work was partially supported by grants from the JHU Applied Physics Laboratory, Nanyang Technological University, and the Government of Israel, and an unrestricted gift from \href{https://www.apptek.com/}{Applications Technology (AppTek)} Inc.

\bibliographystyle{IEEEtran}

\bibliography{mybib}

% Generated by IEEEtran.bst, version: 1.13 (2008/09/30)
\begin{thebibliography}{10}
\providecommand{\url}[1]{#1}
\csname url@samestyle\endcsname
\providecommand{\newblock}{\relax}
\providecommand{\bibinfo}[2]{#2}
\providecommand{\BIBentrySTDinterwordspacing}{\spaceskip=0pt\relax}
\providecommand{\BIBentryALTinterwordstretchfactor}{4}
\providecommand{\BIBentryALTinterwordspacing}{\spaceskip=\fontdimen2\font plus
\BIBentryALTinterwordstretchfactor\fontdimen3\font minus
  \fontdimen4\font\relax}
\providecommand{\BIBforeignlanguage}[2]{{%
\expandafter\ifx\csname l@#1\endcsname\relax
\typeout{** WARNING: IEEEtran.bst: No hyphenation pattern has been}%
\typeout{** loaded for the language `#1'. Using the pattern for}%
\typeout{** the default language instead.}%
\else
\language=\csname l@#1\endcsname
\fi
#2}}
\providecommand{\BIBdecl}{\relax}
\BIBdecl

\bibitem{Hain2012TranscribingMW}
T.~Hain, L.~Burget, J.~Dines, P.~N. Garner, F.~Gr{\'e}zl, A.~E. Hannani,
  M.~Huijbregts, M.~Karafi{\'a}t, M.~Lincoln, and V.~Wan, ``Transcribing
  meetings with the {AMIDA} systems,'' \emph{IEEE Transactions on Audio,
  Speech, and Language Processing}, vol.~20, pp. 486--498, 2012.

\bibitem{Hori2012LowLatencyRM}
T.~Hori, S.~Araki, T.~Yoshioka, M.~Fujimoto, S.~Watanabe, T.~Oba, A.~Ogawa,
  K.~Otsuka, D.~Mikami, K.~Kinoshita, T.~Nakatani, A.~Nakamura, and J.~Yamato,
  ``Low-latency real-time meeting recognition and understanding using distant
  microphones and omni-directional camera,'' \emph{IEEE Transactions on Audio,
  Speech, and Language Processing}, vol.~20, pp. 499--513, 2012.

\bibitem{Renals2017DistantSR}
S.~Renals and P.~Swietojanski, ``Distant speech recognition experiments using
  the {AMI} corpus,'' in \emph{New Era for Robust Speech Recognition,
  Exploiting Deep Learning}, 2017.

\bibitem{Yoshioka2018RecognizingOS}
T.~Yoshioka, H.~Erdogan, Z.~Chen, X.~Xiao, and F.~Alleva, ``Recognizing
  overlapped speech in meetings: A multichannel separation approach using
  neural networks,'' in \emph{INTERSPEECH}, 2018.

\bibitem{barker2018fifth}
J.~Barker, S.~Watanabe, E.~Vincent, and J.~Trmal, ``The fifth {'CHiME'} speech
  separation and recognition challenge: dataset, task and baselines,''
  \emph{arXiv preprint arXiv:1803.10609}, 2018.

\bibitem{Xiong2016AchievingHP}
W.~Xiong, J.~Droppo, X.~Huang, F.~Seide, M.~Seltzer, A.~Stolcke, D.~Yu, and
  G.~Zweig, ``Achieving human parity in conversational speech recognition,''
  \emph{ArXiv}, vol. abs/1610.05256, 2016.

\bibitem{wang2019espresso}
Y.~Wang, T.~Chen, H.~Xu, S.~Ding, H.~Lv, Y.~Shao, N.~Peng, L.~Xie, S.~Watanabe,
  and S.~Khudanpur, ``Espresso: A fast end-to-end neural speech recognition
  toolkit,'' \emph{arXiv preprint arXiv:1909.08723}, 2019.

\bibitem{Lscher2019RWTHAS}
C.~L{\"u}scher, E.~Beck, K.~Irie, M.~Kitza, W.~Michel, A.~Zeyer,
  R.~Schl{\"u}ter, and H.~Ney, ``{RWTH} {ASR} systems for {L}ibri{S}peech:
  Hybrid vs attention - w/o data augmentation,'' in \emph{INTERSPEECH}, 2019.

\bibitem{Synnaeve2019EndtoendAF}
G.~Synnaeve, Q.~Xu, J.~Kahn, E.~Grave, T.~Likhomanenko, V.~Pratap, A.~Sriram,
  V.~Liptchinsky, and R.~Collobert, ``End-to-end {ASR}: from supervised to
  semi-supervised learning with modern architectures,'' \emph{ArXiv}, vol.
  abs/1911.08460, 2019.

\bibitem{zorilatoshiba}
C.~Zorila, M.~Li, D.~Hayakawa, M.~Liu, N.~Ding, and R.~Doddipatla,
  ``Toshiba’s speech recognition system for the {CHiME} 2020 challenge.''

\bibitem{kanda2018hitachi}
N.~Kanda, R.~Ikeshita, S.~Horiguchi, Y.~Fujita, K.~Nagamatsu, X.~Wang,
  V.~Manohar, N.~E.~Y. Soplin, M.~Maciejewski, S.-J. Chen \emph{et~al.}, ``The
  {H}itachi/{JHU} {CHiME-5} system: Advances in speech recognition for everyday
  home environments using multiple microphone arrays,'' in \emph{The 5th
  International Workshop on Speech Processing in Everyday Environments (CHiME
  2018), Interspeech}, 2018.

\bibitem{zorila2019investigation}
C.~Zorila, C.~Boeddeker, R.~Doddipatla, and R.~Haeb-Umbach, ``An investigation
  into the effectiveness of enhancement in {ASR} training and test for
  {CHiME-5} dinner party transcription,'' \emph{arXiv preprint
  arXiv:1909.12208}, 2019.

\bibitem{boeddeker2018front}
C.~Boeddeker, J.~Heitkaemper, J.~Schmalenstroeer, L.~Drude, J.~Heymann, and
  R.~Haeb-Umbach, ``Front-end processing for the {CHiME-5} dinner party
  scenario,'' in \emph{CHiME5 Workshop, Hyderabad, India}, 2018.

\bibitem{ko2017study}
T.~Ko, V.~Peddinti, D.~Povey, M.~L. Seltzer, and S.~Khudanpur, ``A study on
  data augmentation of reverberant speech for robust speech recognition,'' in
  \emph{2017 IEEE International Conference on Acoustics, Speech and Signal
  Processing (ICASSP)}.\hskip 1em plus 0.5em minus 0.4em\relax IEEE, 2017, pp.
  5220--5224.

\bibitem{wang2019jhu}
Y.~Wang, D.~Snyder, H.~Xu, V.~Manohar, P.~S. Nidadavolu, D.~Povey, and
  S.~Khudanpur, ``The {JHU} {ASR} system for {VOiCES} from a {D}istance
  challenge 2019,'' \emph{Proc. Interspeech 2019}, pp. 2488--2492, 2019.

\bibitem{park2019specaugment}
D.~S. Park, W.~Chan, Y.~Zhang, C.-C. Chiu, B.~Zoph, E.~D. Cubuk, and Q.~V. Le,
  ``Specaugment: A simple data augmentation method for automatic speech
  recognition,'' \emph{arXiv preprint arXiv:1904.08779}, 2019.

\bibitem{Zhou2020TheRA}
W.~Zhou, W.~Michel, K.~Irie, M.~Kitza, R.~Schl{\"u}ter, and H.~Ney, ``The
  {RWTH} {ASR} system for {TED-LIUM} release 2: Improving hybrid {HMM} with
  {S}pec{A}ugment,'' \emph{ArXiv}, vol. abs/2004.00960, 2020.

\bibitem{seltzer2013investigation}
M.~L. Seltzer, D.~Yu, and Y.~Wang, ``An investigation of deep neural networks
  for noise robust speech recognition,'' in \emph{2013 IEEE international
  conference on acoustics, speech and signal processing}.\hskip 1em plus 0.5em
  minus 0.4em\relax IEEE, 2013, pp. 7398--7402.

\bibitem{saon2013speaker}
G.~Saon, H.~Soltau, D.~Nahamoo, and M.~Picheny, ``Speaker adaptation of neural
  network acoustic models using i-vectors,'' in \emph{2013 IEEE Workshop on
  Automatic Speech Recognition and Understanding}.\hskip 1em plus 0.5em minus
  0.4em\relax IEEE, 2013, pp. 55--59.

\bibitem{drude2018nara}
L.~Drude, J.~Heymann, C.~Boeddeker, and R.~Haeb-Umbach, ``{NARA-WPE}: A python
  package for weighted prediction error dereverberation in {N}umpy and
  {T}ensorflow for online and offline processing,'' in \emph{Speech
  Communication; 13th ITG-Symposium}.\hskip 1em plus 0.5em minus 0.4em\relax
  VDE, 2018, pp. 1--5.

\bibitem{anguera2007acoustic}
X.~Anguera, C.~Wooters, and J.~Hernando, ``Acoustic beamforming for speaker
  diarization of meetings,'' \emph{IEEE Transactions on Audio, Speech, and
  Language Processing}, vol.~15, no.~7, pp. 2011--2022, 2007.

\bibitem{Nakatani2019AUC}
T.~Nakatani and K.~Kinoshita, ``A unified convolutional beamformer for
  simultaneous denoising and dereverberation,'' \emph{IEEE Signal Processing
  Letters}, vol.~26, pp. 903--907, 2019.

\bibitem{Garca2019SpeakerDI}
P.~Garc{\'i}a, J.~Villalba, H.~Bredin, J.~Du, D.~Cast{\'a}n, A.~Cristia,
  L.~Bullock, L.~Guo, K.~Okabe, P.~S. Nidadavolu, S.~Kataria, S.~Chen,
  L.~Galmant, M.~Lavechin, L.~Sun, M.-P. Gill, B.~Ben-Yair, S.~Abdoli, X.~Wang,
  W.~Bouaziz, H.~Titeux, E.~Dupoux, K.~A. Lee, and N.~Dehak, ``Speaker
  detection in the wild: Lessons learned from {JSALT} 2019,'' \emph{ArXiv},
  vol. abs/1912.00938, 2019.

\bibitem{Sun2020ProgressiveMN}
L.~Sun, J.~Du, X.~Zhang, T.~Gao, X.~Fang, and C.-H. Lee, ``Progressive
  multi-target network based speech enhancement with {SNR}-preselection for
  robust speaker diarization,'' in \emph{ICASSP 2020}, 2020.

\bibitem{Kataria2020FeatureEW}
S.~Kataria, P.~S. Nidadavolu, J.~Villalba, N.~Chen, P.~Garc{\'i}a, and
  N.~Dehak, ``Feature enhancement with deep feature losses for speaker
  verification,'' \emph{ArXiv}, vol. abs/1910.11905, 2020.

\bibitem{bullock2019overlap}
L.~Bullock, H.~Bredin, and L.~P. Garcia-Perera, ``Overlap-aware diarization:
  resegmentation using neural end-to-end overlapped speech detection,''
  \emph{arXiv preprint arXiv:1910.11646}, 2019.

\bibitem{watanabe2020chime}
S.~Watanabe, M.~Mandel, J.~Barker, and E.~Vincent, ``{CHiME-6} challenge:
  Tackling multispeaker speech recognition for unsegmented recordings,''
  \emph{arXiv preprint arXiv:2004.09249}, 2020.

\bibitem{nagrani2017voxceleb}
A.~Nagrani, J.~S. Chung, and A.~Zisserman, ``Voxceleb: a large-scale speaker
  identification dataset,'' \emph{arXiv preprint arXiv:1706.08612}, 2017.

\bibitem{nakatani2010speech}
T.~Nakatani, T.~Yoshioka, K.~Kinoshita, M.~Miyoshi, and B.-H. Juang, ``Speech
  dereverberation based on variance-normalized delayed linear prediction,''
  \emph{IEEE Transactions on Audio, Speech, and Language Processing}, vol.~18,
  no.~7, pp. 1717--1731, 2010.

\bibitem{kanda2019guided}
N.~Kanda, C.~Boeddeker, J.~Heitkaemper, Y.~Fujita, S.~Horiguchi, K.~Nagamatsu,
  and R.~Haeb-Umbach, ``Guided source separation meets a strong {ASR} backend:
  {H}itachi/{P}aderborn university joint investigation for dinner party
  {ASR},'' \emph{arXiv preprint arXiv:1905.12230}, 2019.

\bibitem{Peddinti2015ATD}
V.~Peddinti, D.~Povey, and S.~Khudanpur, ``A time delay neural network
  architecture for efficient modeling of long temporal contexts,'' in
  \emph{INTERSPEECH}, 2015.

\bibitem{sell2018diarization}
G.~Sell, D.~Snyder, A.~McCree, D.~Garcia-Romero, J.~Villalba, M.~Maciejewski,
  V.~Manohar, N.~Dehak, D.~Povey, S.~Watanabe \emph{et~al.}, ``Diarization is
  hard: Some experiences and lessons learned for the {JHU} team in the
  inaugural {DIHARD} challenge.'' in \emph{Interspeech}, vol. 2018, 2018, pp.
  2808--2812.

\bibitem{landini2019but}
F.~Landini, S.~Wang, M.~Diez, L.~Burget, P.~Mat{\v{e}}jka,
  K.~{\v{Z}}mol{\'\i}kov{\'a}, L.~Mo{\v{s}}ner, O.~Plchot, O.~Novotn{\`y},
  H.~Zeinali \emph{et~al.}, ``{BUT} system description for {DIHARD} speech
  diarization challenge 2019,'' \emph{arXiv preprint arXiv:1910.08847}, 2019.

\bibitem{snyder2018x}
D.~Snyder, D.~Garcia-Romero, G.~Sell, D.~Povey, and S.~Khudanpur, ``X-vectors:
  Robust {DNN} embeddings for speaker recognition,'' in \emph{2018 IEEE
  International Conference on Acoustics, Speech and Signal Processing
  (ICASSP)}.\hskip 1em plus 0.5em minus 0.4em\relax IEEE, 2018, pp. 5329--5333.

\bibitem{Dez2018SpeakerDB}
M.~D{\'i}ez, L.~Burget, and P.~Matejka, ``Speaker diarization based on
  {Bayesian} {HMM} with eigenvoice priors,'' in \emph{Odyssey}, 2018.

\bibitem{Ravanelli2018SpeakerRF}
M.~Ravanelli and Y.~Bengio, ``Speaker recognition from raw waveform with
  {S}inc{N}et,'' \emph{2018 IEEE Spoken Language Technology Workshop (SLT)},
  pp. 1021--1028, 2018.

\bibitem{PoveyEtAlIS2018}
\BIBentryALTinterwordspacing
D.~Povey, G.~Cheng, Y.~Wang, K.~Li, H.~Xu, M.~Yarmohammadi, and S.~Khudanpur,
  ``Semi-orthogonal low-rank matrix factorization for deep neural networks,''
  in \emph{Proc. Interspeech 2018}, 2018, pp. 3743--3747. [Online]. Available:
  \url{http://dx.doi.org/10.21437/Interspeech.2018-1417}
\BIBentrySTDinterwordspacing

\bibitem{PoveyEtAlIS2016}
\BIBentryALTinterwordspacing
D.~Povey, V.~Peddinti, D.~Galvez, P.~Ghahremani, V.~Manohar, X.~Na, Y.~Wang,
  and S.~Khudanpur, ``Purely sequence-trained neural networks for asr based on
  lattice-free mmi,'' in \emph{Interspeech 2016}, 2016, pp. 2751--2755.
  [Online]. Available: \url{http://dx.doi.org/10.21437/Interspeech.2016-595}
\BIBentrySTDinterwordspacing

\bibitem{ChenEtAlIS2015}
G.~Chen, H.~Xu, M.~Wu, D.~Povey, and S.~Khudanpur, ``Pronunciation and silence
  probability modeling for asr,'' in \emph{Sixteenth Annual Conference of the
  International Speech Communication Association}, 2015.

\bibitem{stolcke2002srilm}
A.~Stolcke, ``{SRILM}-an extensible language modeling toolkit,'' in
  \emph{Seventh International Conference on Spoken Language Processing}, 2002.

\bibitem{xu2018neural}
H.~Xu, K.~Li, Y.~Wang, J.~Wang, S.~Kang, X.~Chen, D.~Povey, and S.~Khudanpur,
  ``Neural network language modeling with letter-based features and importance
  sampling,'' in \emph{2018 IEEE international conference on acoustics, speech
  and signal processing (ICASSP)}.\hskip 1em plus 0.5em minus 0.4em\relax IEEE,
  2018, pp. 6109--6113.

\bibitem{xu2018pruned}
H.~Xu, T.~Chen, D.~Gao, Y.~Wang, K.~Li, N.~Goel, Y.~Carmiel, D.~Povey, and
  S.~Khudanpur, ``A pruned {RNNLM} lattice-rescoring algorithm for automatic
  speech recognition,'' in \emph{2018 IEEE International Conference on
  Acoustics, Speech and Signal Processing (ICASSP)}.\hskip 1em plus 0.5em minus
  0.4em\relax IEEE, 2018, pp. 5929--5933.

\bibitem{xu2011minimum}
H.~Xu, D.~Povey, L.~Mangu, and J.~Zhu, ``Minimum bayes risk decoding and system
  combination based on a recursion for edit distance,'' \emph{Computer Speech
  \& Language}, vol.~25, no.~4, pp. 802--828, 2011.

\end{thebibliography}

\end{document}